\documentclass{article}

\usepackage{graphicx}
\usepackage{rotating} 
\usepackage{fullpage,amsmath,amssymb,latexsym}
\usepackage{multirow}

\begin{document}
\title{Jupiter's Occultation Radii: Implications for its Internal Dynamics}
\author{ Ravit Helled$^{}$\\
\small{Department of Earth and Space Sciences}\\
\small{University of California, Los Angeles, CA 90095-1567, USA}\\
}

\date{}
\maketitle 

\begin{abstract}
The physical shape of a giant planet can reveal important information about its centrifugal potential, and therefore, its rotation. 
In this paper I investigate the response of Jupiter's shape to differential rotation on cylinders of various cylindrical radii using a simple equipotential theory.  I find that both solid-body rotation (with System III rotation rate) and differential rotation on cylinders up to a latitude between 20 and 30 degrees are consistent with Jupiter's measured shape. Occultation measurements of Jupiter's shape could provide an independent method to constrain the depth of its zonal winds. 
\end{abstract}

\section{Introduction}
Despite decades of investigation, Jupiter's atmospheric circulation is still not well understood. Major questions revolve around Jupiter's dynamical structure, in particular, how its zonal winds are forced (from below by internal heat or from above by solar heating) and how deeply they extend into the interior. 
Many authors have attempted to better constrain and understand Jupiter's dynamics based on various approaches (typically deep convection models vs. shallow atmospheric models) and different numerical methods (e.g., Williams, 1978; Cho and Polvani, 1996; Nozawa and Yoden, 1997; Huang and Robinson, 1998; Williams, 2003; Aurnou and Heimpel, 2004; Heimpel et al., 2005; Vasavada and Showman, 2005; Showman et al., 2006; Heimpel and Aurnou, 2007; Lian and Showman, 2008).\par

Deep convection models of the Jovian zonal winds (e.g., Busse, 1976; Heimpel et al., 2005; Heimpel and Aurnou, 2007) suggest that the bottom boundary of the deep convection is at about 0.9 R$_{\text J}$ (R$_{\text J}$ is the radius of Jupiter). Liu et al. (2008) discuss the importance of Lorentz forces in constraining the depth of the Jovian circulation. They argue that the observed zonal winds on Jupiter cannot extend more deeply than 0.96 R$_{\text J}$. Shallow forcing models typically argue that the zonal winds are driven by moist convection and are a surface phenomenon, i.e., a 'weather layer' (e.g., Ingersoll et al., 1969; Ingersoll et al., 2004). \par

The penetration depth of Jupiter's zonal winds also has implications for the planet's internal structure. If the zonal winds are 'deep' and involve a non-negligible amount of mass, interior models that typically assume solid-body (SB) rotation must be modified, and corrections to the gravitational coefficients due to dynamics must be included (Hubbard, 1982; 1999). A recent interior model of Jupiter by Militzer et al.~(2008) suggests that Jupiter's measured $J_4$ value can only be fit when differential rotation is considered because no interior models that fit Jupiter's $J_4$ could be found under the assumption of solid-body rotation. The interior model presented by Militzer et al.~(2008), however, consists of 2-layers (a heavy-element core surrounded by a hydrogen-helium envelope), while standard 3-layer Jupiter models with solid-sbody rotation can typically fit Jupiter's gravitational coefficient ($J_2, J_4, J_6$) (e.g., Saumon \& Guillot, 2004; Nettelmann et al., 2008). 
As the question of whether Jupiter's interior structure is consistent with solid-body rotation remains open, it is clear that a better determination of the depth of Jupiter's zonal winds is desirable for both dynamical and interior models. \par

Jupiter's internal dynamics, or more precisely, its rotation profile, can be constrained by measurements of its high-order gravitational coefficients. Hubbard and collaborators (e.g., Hubbard, 1982; 1999; Kaspi, et al., 2010) showed that a departure from solid-body rotation is detectable in gravitational coefficients larger than about degree 10 (Kaspi et al., 2010, figure 7b). The Juno mission to Jupiter (Bolton, 2005) is designed to provide accurate measurements of Jupiter's gravitational field to high order and can therefore constrain its dynamical structure. Gravitational anomalies can also be used directly to infer Jupiter's density anomalies and therefore its rotation profile (see Kaspi et al., 2010 and references therein). In this paper I use an equipotential theory to derive Jupiter's shape when differential rotation on cylinders is included and suggest that shape data, i.e., occultation radii, can provide an independent method of constraining its internal rotation profile. 

\section{Jupiter's Geoid with Differential Rotation on Cylinders}

The effective potential $U$ of a giant planet in hydrostatic equilibrium is given by a combination of its gravitational potential and its centrifugal potential (Kaula, 1968), 
\begin{equation}
U = \frac{G M}{r} \left( 1 - \sum_{n=1}^\infty \left( \frac{a}{r} \right)^{2 n} J_{2 n} P_{2 n} \left( \mu  \right)  \right)+ \frac{1}{2} q \left( \frac{G M}{a} \right) \left( \frac{r}{a} \right)^2 (1-\mu^2).
\label{U}
\end{equation}
where $G M$ is the gravitational constant times the total planetary mass, $a$ is the equatorial radius, $J_{2 n}$ are the gravitational coefficients, and $P_{2n}$ are the Legendre polynomials.  The point where the potential is evaluated is given by the planetocentric radius $r$ and latitude $\phi$ ($\mu=\text{sin}\phi$).
The second term on the right side is the centrifugal potential. $q=\omega^2a^3$/$GM$ is the smallness parameter, where $\omega$ is the angular velocity of rotation. The planet can rotate uniformly as a solid-body, or on cylinders to a certain depth, including differential rotation on cylinders all the way to the planet's center. The harmonic coefficients $J_{2n}$ are obtained from the measured values for a reference equatorial radius of 71,492 km (Jacobson, 2003). The values of $J_{2n}$ from Jacobson (2003) in units of $10^{-6}$ are $J_2$ = 14696.43 $\pm$0.21, $J_4$ = -587.14 $\pm$ 1.68, $J_6$ = 34.25 $\pm$ 5.22. \par 

Helled et al. (2009) discuss how the planetary shape derived from the planet's effective potential, can be used to gain information on its internal rotation. Helled et al. (2009)  searched for a solid-body rotation period that best fits Jupiter's measured physical shape, i.e., a solid-body rotation period that minimizes the planet's dynamical heights (for details, see Helled et al. 2009 and Lindal, 1992). The physical shape of Jupiter was derived by adding the dynamical heights (obtained using wind data) to the calculated geoid. The shape of Jupiter's geoid in Helled et al.~(2009) was derived assuming solid-body rotation, including a range of rotation rates. Since Jupiter may not rotate solely as a solid-body, it is desirable to perform an investigation of Jupiter's geoid shape in which the possibility of differential rotation is included. In the analysis below I project Jupiter's zonal winds along cylinders to describe its rotation profile, which is explicitly used in the reference effective potential equation (Eq.~1). The derived shape using different cylindrical radii is then compared with Jupiter's measured shape. Figure 1 shows Jupiter's physical shape as calculated by Helled et al.~(2009) with six measured occultation radii from the Pioneer and Voyager missions (see Fig. 7 in Lindal et al., (1981).  \par

When differential rotation is included, the planetary rotation rate is no longer given by a constant ($\omega_0$); instead, the rotation profile changes with cylindrical radius. 
I assume that Jupiter rotates on cylinders about the polar axis, so the angular velocity profile $\omega$ depends on the distance of a cylindrical shell from the spin axis (Helled et al., 2010). I define $\gamma$ as the normalized cylindrical polar radius, and derive the angular velocity $\omega(\gamma)$ using Jupiter's wind data. 
I use Jupiter's zonal wind velocities obtained from Hubble Space Telescope images from 1995 to 2000 (Garc\'{i}a-Melendo \& S\'{a}nchez-Lavega, 2001) at three different wavelengths (410, 892, 953 nm) representing Jupiter's cloud top levels ($\sim$ 1 bar). 
The value of $\gamma$ ranges between zero and one; $\gamma=1$ corresponds to solid-body rotation, and $\gamma=0$ refers to differential rotation on cylinders all the way to Jupiter's center.  Figure 2 shows a sketch of the planet's rotation profile for a given normalized cylindrical polar radius value ($\gamma_0$). The planet rotates as a solid-body (with a rotation rate $\omega_0$) up to $\gamma_0$,  from that point on the planet rotates differentially on cylinders with a rotation rate profile $\omega(\gamma)$. $\phi_{\omega_{diff}}$ is the latitude corresponding to $\gamma_0$ that defines the latitude in which transition between solid-body and differential rotation occurs.  
 \par

Jupiter's zonal wind velocities are available for both the northern and southern hemispheres. Although the wind profiles are similar in the two hemispheres, they are not identical.  Figure 3 shows $\omega(\gamma)$ using the northern (dashed black curve) and southern (dashed gray curve) wind velocities separately. The horizontal thin black line shows the value of $\omega_0$, Jupiter's angular velocity for a solid-body rotation period of 9h 55m 29.56s (Higgins et al., 1996). 
Note that $\gamma=0$ corresponds to the planet's pole ($\phi_{\omega_{diff}}$=90$^o$, i.e., differential rotation on cylinders all the way to the center) while the wind velocity data do not reach the poles. In the northern hemisphere, wind velocities are measured to higher latitudes and therefore can reach smaller $\gamma$ values. The thick black curve shows the angular velocity averaged over the northern (dashed black curve) and southern (dashed gray curve) hemispheres. 
I can now compute Jupiter's centrifugal potential for different values of $\gamma$, i.e., differential rotation on cylinders for different cylindrical polar radii. \par

Reference geoids representing Jupiter's shape (Helled et al., 2009), are defined in the usual manner as the level surfaces of equal effective potential, and are derived for both the 100 mbar and 1 bar pressure-levels. The advantage in using the 100 mbar level is that radio occultation data refer directly to this pressure-level. The zonal wind velocities, however,  correspond to $\sim$ 1 bar. As shown below, I get similar results for both pressure-levels. Jupiter's polar radius is held fixed to its measured value and is used to define the equipotential surface. The polar radii for the 100 mbar and 1 bar pressure levels are 66,896$\pm$4 km and 66,854$\pm$4 km, respectively (Lindal et al., 1981; Lindal, 1992). The geoid's  radius is calculated to fifth order; the harmonics $J_2$, $J_4$, and $J_6$ are extrapolated to even harmonics $J_8$ and $J_{10}$ to ensure that the precision of the geoid calculation is better than $\pm$ 1.0 km. There is an implicit assumption in the procedure that the mass in the atmosphere above the calculated pressure level is negligible. The computation therefore assumes that non-hydrostatic effects have a negligible contribution to the planetary shape. More details on the method can be found in Helled et al.~(2009, 2010) and references therein. 
\par

\section{Results} 
The six radio occultation radii of Jupiter's 100 mbar isosurface at different planetocentric latitudes, four of them at the northern hemisphere, are shown in Fig.~1. The formal error on Jupiter's measured radii is $\sim$4 km (Lindal et al, 1981; Lindal, 1992), and can therefore be taken as the standard error bar on Jupiter's shape. Helled et al. (2009) presented the extrapolated occultation radii at the 1 bar pressure level by subtracting 47.82 km along the vertical to the geoid and
projecting it on the geocentric radius. The estimate of the 47.82 km height
of the 100 mbar level above the one-bar level in the atmosphere is obtained
by interpolating in Table 8.2 in Lodders \& Fegley (1998).  
\par

Below I vary the differential rotation polar cylindrical radius and investigate how Jupiter's shape is changing accordingly. I compute Jupiter's shape assuming that differential rotation on cylinders occurs up to each available occultation measurement at a given planetocentric latitude $\phi$. I then compare the calculated shape with the measurements. The planetary shape is adjusted so the radii at the transition between differential and solid-body rotation converge (Helled et al., 2010). I.e., for planetocentric latitudes greater than $\phi_{\omega_{diff}}$ the planetary shape is similar to that of Jupiter with a solid-body rotation. Although this assumption ensures that the planetary shape is an equipotential surface, it neglects the zonal winds at latitudes larger than $\phi_{\omega_{diff}}$. As a result, configurations in which differential rotation penetrates only partially at high latitudes (e.g., Hiempel et al., 2005) are not considered. The geostrophic balance on jets at high latitudes would force the 100 mbar (or 1 bar) pressure-level to exhibit topography relative to the equipotential, and therefore solid-body rotation would lead to smaller radii at these latitudes. However, the amplitude of this effect is $\sim$ 1-2 km, smaller than the formal error bar on Jupiter's radii. 

\subsection{Averaging North and South}  
Table 1 shows the calculated radii for the occultation latitudes at both the 100 mbar (top) and 1 bar (bottom) pressure-levels when the zonal winds are averaged between north and south. $r_{occ}$ is the measured radius at a given pressure-level. $\Delta r$ of SB rotation gives the radii residuals in km for solid-body rotation. $\Delta r$ of $\omega_{diff}$ at a given planetocentric latitude is the difference (in km) between the measured and computed radii assuming differential rotation up to a given $\phi_{\omega_{diff}}$ (see figure 2). 
For latitudes $\>\phi_{\omega_{diff}}$ the residuals are similar to those computed for SB rotation. \par
Since the wind velocities are averaged between the northern and southern hemispheres, results are presented for all occultation latitudes, and for both 100 mbar and 1 bar pressure levels. Generally, better agreement with the measurements, regardless of the rotation profile, is found for the 100 mbar pressure-level. This is due to the lack of extrapolation in radius as done for the 1 bar pressure-level (Helled et al., 2009). The residuals found here are typically larger than the ones presented by Helled et al. (2009) because I do not include the offset in north-south symmetry, which is found to be $\sim 5.5$ km for Jupiter (for details, see Helled et al., 2009 and references therein).
For both pressure-levels, solid-body rotation leads to shapes that agree well with occultation data, as found by Helled et al., (2009). Differential rotation on cylinders, depending on its cylindrical polar radius ('depth'), could lead to similar agreement as long as differential rotation does not penetrate all the way to Jupiter's center. Given that the error on Jupiter's shape is $\sim$4 km, residuals in radius up to about 5 km can be considered as a good fit. In this case the results suggest that differential rotation on cylinders above $\phi_{\omega_{diff}} \sim 20^o$ can be ruled out for Jupiter. It should be noted that when differential rotation up to higher latitudes is considered, the discrepancy with observations occurs at low-latitude regions. Differential rotation at high latitudes can lead to agreement with high-latitude shape but not with the equatorial shape. \par

In order to better understand this behavior, one could consider the following exercise. As seen in Fig. 3, Jupiter's jets are mostly eastward, and thus a model with jets that penetrate through the planet essentially has a slightly faster rotation rate than a planet in solid-body rotation at the System III rotation rate. A typical speed of the mid- and high-latitude jets is $\sim$ 20 m s$^{-1}$, about 0.2\% of the speed of the planet's rotation in inertial  space. To zero-order, increasing the rotation rate by 0.2\% should increase the rotational flattening by the same factor. The difference in equatorial and polar radii for Jupiter is about 4600 km. Thus, increasing the flattening by 0.2\%, would imply an increase in the equatorial radius (relative to the polar radius) of the order $4600\times0.002 = 9$ km. In comparison, Table 1 quotes equatorial radii that are about 11 km larger when differential rotation up to high latitudes is included.

\subsection{Separating North and South}
I repeat the calculation, this time separating the wind data for the northern and southern hemispheres. The results are summarized in Table 2. For the northern hemisphere, the differences between the measured radii and the calculated ones are slightly worse than in the previous case; the results, however, remain the same. Again, I find that solid-body rotation provides a good fit to wind data as well as differential rotation up to a latitude $\phi_{\omega_{diff}}\sim20^o$. Differential rotation toward Jupiter's center results in larger discrepancy with the shape data. \par

For the southern hemisphere, only two measurements are available, one of them is at latitude of 71.8$^o$ S. Since Jupiter's zonal wind velocities in the southern hemisphere are not available at this latitude, only the occultation measurement at 10.1$^o$ S is presented. The results suggest that both solid-body rotation and differential rotation on cylinders up to $\phi_{\omega_{diff}}=10^o$ are consistent with Jupiter's measured shape, a result which is consistent with the previous cases. \par 

\section{Summary}
The centrifugal potential of a giant planet influences its shape. Knowledge of Jupiter's shape can be used to better constrain the depth of its zonal winds. Better determination of the zonal winds' penetration depth is important for the understanding of Jupiter's global circulation, as well as its magnetic field generation (e.g., Ingersoll et al., 2004).  
I present a simple computation of Jupiter's geoid when differential rotation on cylinders is included and compare its shape to available shape data. To keep the planet in hydrostatic equilibrium and treat its shape as an equipotential surface, I define differential rotation on cylinders about the polar axis. For this rotational configuration, I find that the maximum latitude to which differential rotation on cylinders can occur lies between 20 and 30 degrees. \par 

While the work presented here assumes differential rotation on cylinders, which may be a good representation of Jupiter's internal dynamics, other rotational profiles may be possible, such as differential rotation on spheres, as seems to be the case for the Sun (Elliott et al., 2000; Glatzmaier, 2008). 
Although further investigation of the topic is desirable; in particular, shape models that account for non-hydrostatic contributions, and 2-D interior models, the analysis presented here suggests that occultation shape data could be used to improve our understanding of Jupiter's internal dynamics. 

\subsection*{Acknowledgments}
I thank John D. Anderson and Gerald Schubert for fruitful discussions. I thank two anonymous referees for valuable comments and suggestions. 
I also acknowledge support from NASA through the Southwest Research Institute. 

\section{References}
Aurnou, J. M. and M. H. Heimpel, 2004, Zonal jets in rotating convection with mixed mechanical 
boundary conditions. Icarus, 169, 492--498. \\
Bolton, S. (2005). Juno final concept study report. Technical Report AO-03-OSS-03, New Frontiers.\\
Cho, J. Y. K. and L. M. Polvani, 1996, The morphogenesis of bands and zonal winds in the 
atmospheres on the giant outer planets. Science, 273, 335--337. \\
Elliott, J. R., M. S. Miesch, and J. Toomre, 2000, Turbulent solar convection and its coupling 
with rotation: The effect of Prandtl number and thermal boundary conditions on the resulting 
differential rotation. Astrophys. J., 533, 546--556. \\
Garc\'{i}a-Melendo \& S\'{a}nchez-Lavega, 2001, A Study of the Stability of Jovian Zonal Winds from HST Images: 1995-2000. Icarus, 152, 316--330.\\
Glatzmaier, G. A., 2008, A note on ÒConstraints on deep-seated zonal winds inside Jupiter and 
SaturnÓ. Icarus, doi:10.1016/j.icarus.2008.03.018. \\
Higgins, C. A., Carr, T. D. \& Reyes, F. 1996. A new determination of Jupiter's radio rotation period. Geophys. Res. Lett., 23, 2653--2656\\
Heimpel, M. H. and J. M. Aurnou, 2007, Turbulent convection in rapidly rotating spherical 
shells: A model for equatorial and high latitude jets on Jupiter and Saturn. Icarus, 187, 540--557.\\
Heimpel, M. H., J. M. Aurnou, and J. Wicht, 2005, Simulation of equatorial and high-latitude 
jets. Nature, 438, 193--196.\\
Helled R., G. Schubert, and J. D. Anderson, 2009. Jupiter and Saturn rotation periods. Planet. Space
Sci., 57, 1467Ð1473. \\
Helled, R., Anderson, J. D., and Schubert, G., 2010. Uranus and Neptune: Shape and Rotation. Icarus, 210, 446--454. \\
Hubbard, W. B., 1982. Effects of Differential Rotation on the Gravitational Figures of Jupiter and Saturn. Icarus, 52, 509--515.\\
Hubbard, W. B., 1999, NOTE: Gravitational Signature of Jupiter's Deep Zonal Flows. Icarus, 137, 357--359.\\    
Ingersoll, A. P., 1970. Motions in Planetary Atmospheres and the Interpretation of Radio Occultation Data. Icarus, 13, 34\\
Ingersoll, A. P. \& Pollard, D., 1982, Motion in the interiors and atmospheres of Jupiter and Saturn - Scale analysis, anelastic equations, barotropic stability criterion. Icarus, 52, 62--80.\\
Huang, H.-P. and W. A. Robinson, 1998, Two-dimensional turbulence and persistent zonal jets in 
a global barotropic model. J. Atmos. Sci., 55, 611--632. \\
Jacobson, R. A. 2003, JUP230 orbit solution, http://ssd.jpl.NAsa.gov/?gravity\_fields\_op \\
Kaspi, Y., Hubbard, W. B., Showman, A. P. and Flierl, G. R., 2010. Gravitational signature of JupiterÕs internal dynamics. Geophys. Res. Lett., 37, L01,204. \\
Kaula, W. M. 1968, An introduction to planetary physics - The terrestrial planets (Space Science Text Series, 
New York: Wiley, 1968) \\
Lindal, G. F. 1992, The atmosphere of Neptune - an analysis of radio occultation data acquired with Voyager 2. AJ, 103, 967--982.\\
Lindal, G. F., Sweetnam, D. N., \& Eshleman, V. R. 1985, The atmosphere of Saturn - an analysis of the Voyager radio occultation measurements. AJ, 90, 1136--1146.\\
Lindal, G. F., Wood, G. E. , Levy, G. S., Anderson, J. D. , Sweetnam, D. N., 
Hotz, H. B., Buckles, B. J., Holmes, D. P. , Doms, P. E. , Eshleman, V. R., 
Tyler, G. L. \& Croft, T. A., 1981. The atmosphere of Jupiter - an analysis of the
Voyager radio occultation measurements. J. Geophys. Res., 86, 8721--8727.\\
Lian, Y. and A. P. Showman, 2008, Deep jets on gas-giant planets. Icarus, 194, 597--615.\\
Liu, J., Goldreich, P. M. and Stevenson, D. J., 2008. Constraints on deep-seated zonal winds inside Jupiter and Saturn. Icarus, 196, 653Ð664.
Lodders, K. \& Fegley, B., 1998. The planetary scientistÕs companion / Katharina
Lodders, Bruce Fegley. The planetary scientistÕs companion / Katharina
Lodders, Bruce Fegley. New York : Oxford University Press, 1998. QB601.L84 1998.\\
Militzer, B., Hubbard, W. B., Vorberger, J., Tamblyn, I. and Bonev, S. A., 2008. A massive core in Jupiter predicted from first-principles simulations. Astrophys. J. Lett., 688, L45ÐL48.\\
Nozawa, T. and S. Yoden, 1997, Formation of zonal band structure in forced two-dimensional 
turbulence on a rotating sphere. Phys. Fluids, 9, 2081--2093. \\
Null, G. W., 1976. Gravity field of Jupiter and its satellites from Pioneer 10 and
Pioneer 11 tracking data. ApJ, 81:1153Ð1161, December 1976.\\
Nettelmann, N., Holst, B., Kietzmann, A., French, M. and Redmer, R. 2008. Ab initio equation of state data for hydrogen, helium, and water and the internal structure of Jupiter. Astrophys. J., 683, 1217Ð1228.\\
Saumon,D. and Guillot,T., 2004. Shock Compression of Deuterium and the Interiors of Jupiter and Saturn. The Astrophysical Journal, 609, 1170--1180.\\
Showman, A. P., P. J. Gierasch, and Y. Lian, 2006, Deep zonal winds can result from shallow 
driving in a giant-planet atmosphere. Icarus, 182, 513--526. \\
Vasavada, A. R. and A. P. Showman, 2005, Jovian atmospheric dynamics: An update after Galileo 
and Cassini. Rep. Prog. Phys., 182, 513--526. \\
Williams, J. G., 1978, Planetary circulations .1. Barotropic representation of Jovian and terrestrial 
turbulence. J. Atmos. Sci., 35, 1399--1426. \\
Williams, J. G., 2003, Jovian dynamics. Part III: Multiple, migrating, and equatorial jets. J. Atmos. 
Sci., 60, 1270--1296. \\
Zharkov, V. N., \& Trubitsyn, V. P. 1978, Physics of planetary interiors (Astronomy and Astrophysics Series,
Tucson: Pachart, 1978) 

\newpage

\begin{figure}[h!]
   \centering
    \includegraphics[width=5.50in]{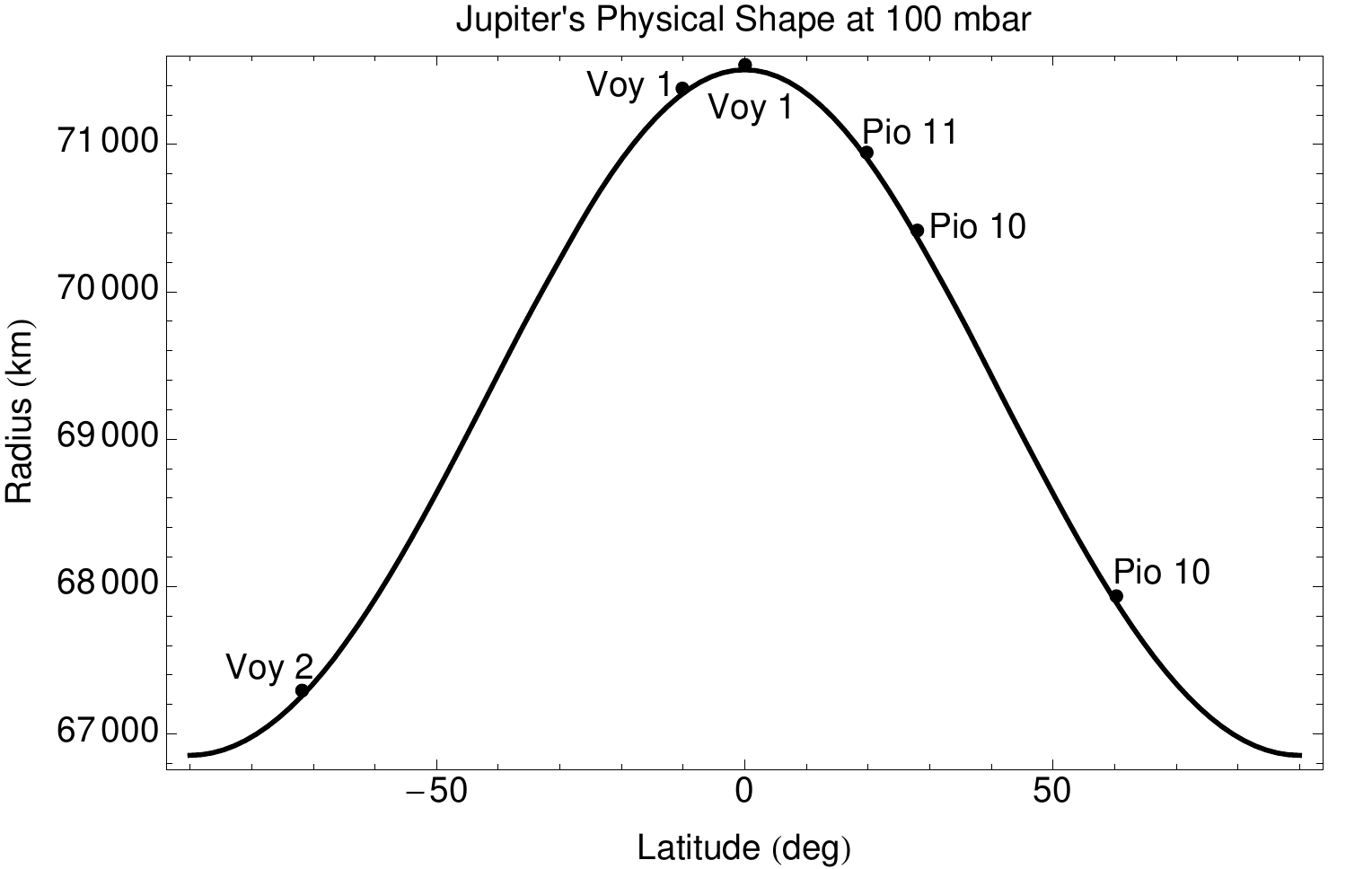}
    \caption[err]{Jupiter's physical shape at 100 mbar (Helled et al., 2009). The curve represents the shape that best fits the radio data. The discrete points correspond to the six occultation radii measured by Pioneer 10 and 11, and Voyager 1 and 2.}
\end{figure}

\begin{figure}[h!]
   \centering
    \includegraphics[width=4.50in]{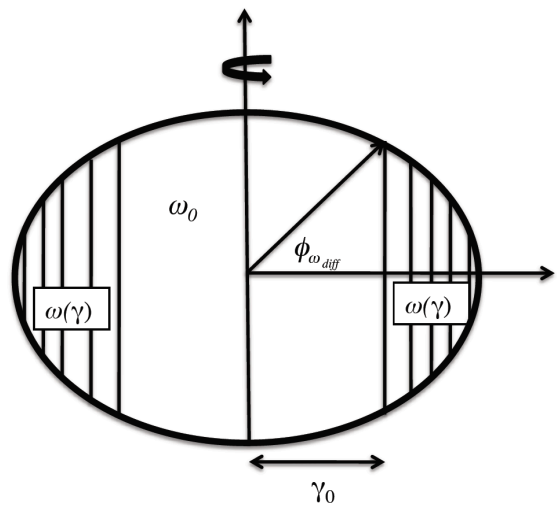}
    \caption[err]{A sketch of Jupiter's rotation profile. $\omega_0$ is the solid-body rotation rate, $\omega(\gamma)$ is the rotation rate assuming differential rotation on cylinders. $\gamma_0$ is the normalized cylindrical radius that defines the boundary between solid-body and differential rotation. $\phi_{\omega_{diff}}$ is the latitude of the transition, and essentially determines the  cylindrical radius of differential rotation as described in the text.}
\end{figure}

\clearpage

\begin{figure}[h!]
   \centering
    \includegraphics[width=5.0in]{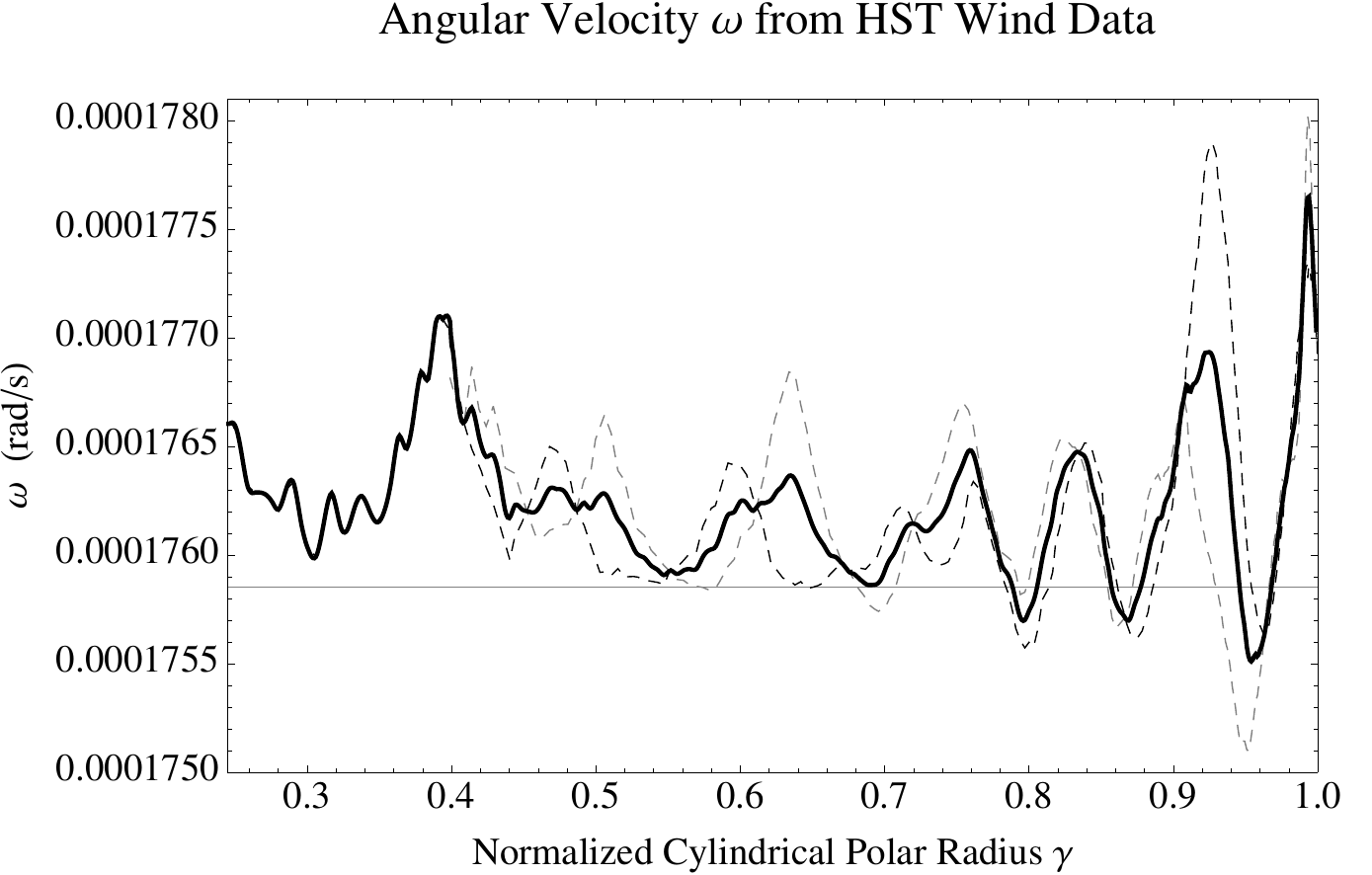}
    \caption[err]{Angular velocity $\omega$ as a function of normalized cylindrical polar radius. The black and gray dashed curves correspond to the northern and southern hemispheres, respectively. The thick black curve shows the average northern and southern values. For $\gamma <0.4$ the thick black curve is identical to the angular velocity of the northern hemisphere since there are no wind data available for high latitudes for Jupiter's southern hemisphere.}
\end{figure}

\newpage
\begin{sidewaystable}[h!]

\caption[Hello]{
	\label{radii}
	Jupiter's shape as calculated from the model for different cylindrical polar radii when Jupiter's zonal winds are averaged over the two hemispheres (see text). The first two columns are Jupiter's planetocentric latitude $\phi$ and radius $r$ obtained by occultation measurements for both hemispheres (see Table 1). The other columns give the difference in radius in km between the measured radius and the calculated radius assuming different rotation profiles. Column three presents the radius residuals assuming solely solid-body (SB) rotation. The rest columns lists the residuals in radius for differential rotation on cylinders up to a given latitude $\phi_{\omega_{diff}}$. Results are shown for both 100 mbar (top) and 1 bar (bottom) pressure-levels. }
\begin{tabular}{lcccccccl}
\hline
\multicolumn{5}{c}{\bf Jupiter's Radii (100 mar) -  Assuming North-South Symmetry} \\
$\phi$  (deg) & $r_{occ}$ (km) & $\Delta r$ (km)& $\Delta r$ (km)&$\Delta r$ (km)&  $\Delta r$ (km) &  $\Delta r$ (km) &  $\Delta r$ (km)&  $\Delta r$ (km)\\
 &&SB rotation  & $\phi_{\omega_{diff}}$= 1$^o$ & $\phi_{\omega_{diff}}$=11$^o$ &   $\phi_{\omega_{diff}}$=20$^o$ &  $\phi_{\omega_{diff}}$=30$^o$&  $\phi_{\omega_{diff}}$=61$^o$&  $\phi_{\omega_{diff}}$=72$^o$ \\
\hline
-71.8 & 67293.64 & 5.38 & SB & SB & SB & SB & SB & 5.42 \\
60.3 & 67933.93 & -4.45& SB & SB & SB & SB & -4.32 & -1.98 \\
28.0 & 70415.08 & -6.51 & SB & SB & SB & -6.09& -0.77 & 1.77 \\
19.8 & 70943.95 & -2.11 & SB & SB &-2.06 & 0.24 & 5.66 & 8.24 \\
-10.1 &71378.73 & -1.56 & SB  & -1.41 &0.60 & 2.94 & 8.44 & 11.06\\
0.07 & 71538.61 &  0.58 & 0.58 & 1.68 & 3.69& 6.05 & 11.58 & 14.22 \\

\hline
\multicolumn{5}{c}{\bf Jupiter's Radii (1 bar) -  Assuming North-South Symmetry} \\
\hline
\hline

$\phi$  (deg) & $r_{occ}$ (km) & $\Delta r$ (km)& $\Delta r$ (km)&$\Delta r$ (km)&  $\Delta r$ (km) &  $\Delta r$ (km) &  $\Delta r$ (km)&  $\Delta r$ (km)\\
 &&SB rotation  & $\phi_{\omega_{diff}}$= 1$^o$ & $\phi_{\omega_{diff}}$= 11$^o$ &  $\phi_{\omega_{diff}}$=20$^o$ &  $\phi_{\omega_{diff}}$=30$^o$&  $\phi_{\omega_{diff}}$=61$^o$&  $\phi_{\omega_{diff}}$=72$^o$ \\
\hline
-71.8 & 67245.85 & 11.95 & SB & SB & SB & SB & SB & 11.99 \\
60.3 & 67886.18 & 1.30 & SB & SB & SB & SB & 1.43 & 3.69 \\
28.0 & 70367.34 & -4.46 & SB & SB & SB & -4.06 & 1.36 & 3.80 \\
19.8 & 70896.17 & -0.83 & SB & SB & -0.78 & 1.47 & 6.98 & 9.48 \\
-10.1 & 71330.92 & -1.02 & SB & -0.85 & 1.20 & 3.49 & 9.08 & 11.61 \\
0.07 & 71490.79 &  0.81 & 0.82& 1.97 & 4.04 & 6.33 & 11.96 & 14.51 \\
\hline

\end{tabular}

\end{sidewaystable}

\begin{table}[h!]
\caption[Hello]{
	\label{radii}
	Jupiter's shape as calculated from the model for different cylindrical polar radii with Jupiter's zonal winds for the northern (top) and southern (bottom) hemispheres. The first two columns are Jupiter's planetocentric latitude $\phi$ and radius $r$ obtained by occultation measurements at Jupiter's northern hemisphere. The other columns give the difference in radius in km between the measured radius and the calculated radius assuming different rotation profiles. Column three presents the radius residuals assuming solely solid-body (SB) rotation. The rest columns lists the residuals in radius for differential rotation on cylinders up to a given latitude $\phi_{\omega_{diff}}$. }
\begin{tabular}{lcccccl}
\hline
\multicolumn{5}{c}{\bf Jupiter's Radii (100 mbar)- Northern Hemisphere} \\
\hline
$\phi$  (deg) & $r_{occ}$ (km) & $\Delta r$ (km)& $\Delta r$ (km)&$\Delta r$ (km)&  $\Delta r$ (km) &  $\Delta r$ (km) \\
 &&SB rotation  & $\phi_{\omega_{diff}}$= 1$^o$ & $\phi_{\omega_{diff}}$= 20$^o$ &   $\phi_{\omega_{diff}}$=30$^o$ & $\phi_{\omega_{diff}}$=61$^o$   \\
\hline
60.3 & 67886.18  & -4.45 & SB & SB & SB & -4.32 \\
28.0 & 70367.34  & -6.51 & SB & SB & -6.02 & -2.08\\
19.8 & 70896.17 &  -2.11& SB & -2.04& 0.96 & 4.97 \\
0.07 & 71490.79 &   0.58 & 0.59 & 5.43 & 8.50 & 12.59\\
\hline
\multicolumn{5}{c}{\bf Jupiter's Radii (1 bar)- Northern Hemisphere} \\
\hline
$\phi$  (deg) & $r_{occ}$ (km) & $\Delta r$ (km)& $\Delta r$ (km)&$\Delta r$ (km)&  $\Delta r$ (km) &  $\Delta r$ (km) \\
&&SB rotation  & $\phi_{\omega_{diff}}$= 1$^o$ & $\phi_{\omega_{diff}}$=20$^o$ & $\phi_{\omega_{diff}}$=30$^o$ &  $\phi_{\omega_{diff}}$=61$^o$   \\
\hline
60.3 & 67886.18  & 1.30 & SB & SB& SB & 1.42 \\
28.0 & 70367.34  & -4.46 & SB & SB & -3.99 & 0.04\\
19.8 & 70896.17 & -0.83& SB & -0.76& 2.18 & 6.29 \\
0.07 & 71490.79 & 0.81 & 0.82 & 5.77 & 8.77 & 12.96 
\\
\hline
\multicolumn{4}{c}{\bf Jupiter's Radii (100 mbar)- Southern Hemisphere} \\
\hline
\hline
-10.1 & 71330.92 & -1.56 & -1.52 \\
\hline
\multicolumn{4}{c}{\bf Jupiter's Radii (1 bar) - Southern Hemisphere} \\
\hline
\hline
-10.1 & 71330.92 & -1.02 & -0.98 \\
\hline
\end{tabular}
\end{table}

\end{document}